\documentclass[11pt]{article}

\input epsf
\usepackage{amsmath,amssymb}
\usepackage[dvips]{color}
\usepackage{graphicx}
\usepackage{epsfig}

\newcommand{\be}{\begin{equation}}
\newcommand{\ee}{\end{equation}}
\newcommand{\bea}{\begin{eqnarray}}
\newcommand{\eea}{\end{eqnarray}}
\newcommand{\ba}{\begin{array}}
\newcommand{\ea}{\end{array}}

\def \nn {\nonumber}
\newcommand{\eq}[1]{(\ref{#1})}

\begin{document}

\begin{titlepage}
 
\vfill

\begin{center}
{\hfill RUNHETC-2010-12}

\bigskip

{\Large \bf{Dynamical Instability of Holographic QCD at Finite Density}}

\vfill

Wu-Yen Chuang$^1$, Shou-Huang Dai$^2$, Shoichi Kawamoto$^2$, Feng-Li Lin$^2$, Chen-Pin Yeh$^3$

\bigskip

{\it $^1$ NHETC, Department of Physics, Rutgers University\\}
{\it 126 Frelinghuysen Rd, Piscataway, New Jersey 08854, USA\\}
{\it $^2$ Department of Physics, National Taiwan Normal University, Taipei, Taiwan\\}
{\it $^3$ Department of Physics, National Taiwan University, Taipei, Taiwan\\}

\end{center} 

\vfill
\begin{abstract}
In this paper we study the dynamical instability of Sakai-Sugimoto's
holographic QCD model at finite baryon density.
In this model, the baryon density, represented by the smeared
instanton on the worldvolume of the probe $D8$-$\overline{D8}$ mesonic
brane, sources the worldvolume electric field, and through the
Chern-Simons term
 it will induces the instability to form a chiral helical wave. 
This is similar to Deryagin-Grigoriev-Rubakov instability to form the
chiral density wave for large $N_c$ QCD at finite density. 
Our results show that this kind of instability occurs for sufficiently
high baryon number densities.
The phase diagram of holographic QCD will thus be changed from
the one which is based only on thermodynamics.
This holographic approach provides an effective way to
study the phases of QCD at finite density, where the conventional
perturbative QCD and lattice simulation fail.

\end{abstract}
\vfill

\end{titlepage}

\setcounter{footnote}{0}

\section{Introduction}

The quantum chromodynamics (QCD) at high baryon density has
implications for physics in many fields, such as the possible
existence of quark matters in the core of compact stars. However,
unlike the physics for QCD at finite temperature, it is difficult to
directly probe such a regime theoretically, even from the first
principle numerical simulations. This is mainly due to the complex
fermion determinant at finite density, which causes the sign problem
in simulation.

Despite that, it was conjectured that there is a color superconducting
ground state for QCD at high enough density\cite{Alford:1997zt},
similar to the BCS mechanism for the formation of Cooper pairs of
quarks near the Fermi surface. Since the Fermi momentum plays the role
of an additional energy scale, we can apply the perturbative
techniques in the regime of ultra-high density. However, at moderate
density where the most interesting physics dwells, QCD is strongly
coupled and it is hard to have reliable treatments.

Instead, one can consider QCD at large $N_c$ (number of color) limit
to probe the above non-perturbative regime of finite density. This can
be done either by using the renormalization group improved
perturbative technique for small ' t Hoof coupling, or by studying the
holographic dual gravity for large 't Hooft coupling. For the former
case, Deryagin, Grigoriev and Rubakov (DGR)\cite{DGR} noticed that
color superconductivity will be suppressed in the large $N_c$ limit
since the Cooper pair of quarks is not color singlet and its formation
is diagramtically non-planar. Moreover, they found that there is a new
dynamical instability for the formation of chiral density wave whose
wave number is the twice of the Fermi momentum \cite{DGR,Shuster:1999tn},
i.e., a spatially modulated phase. This can be understood as the
standing wave from the formation of quark-hole pairs.

\begin{figure}[h]
\centering
\epsfig{file=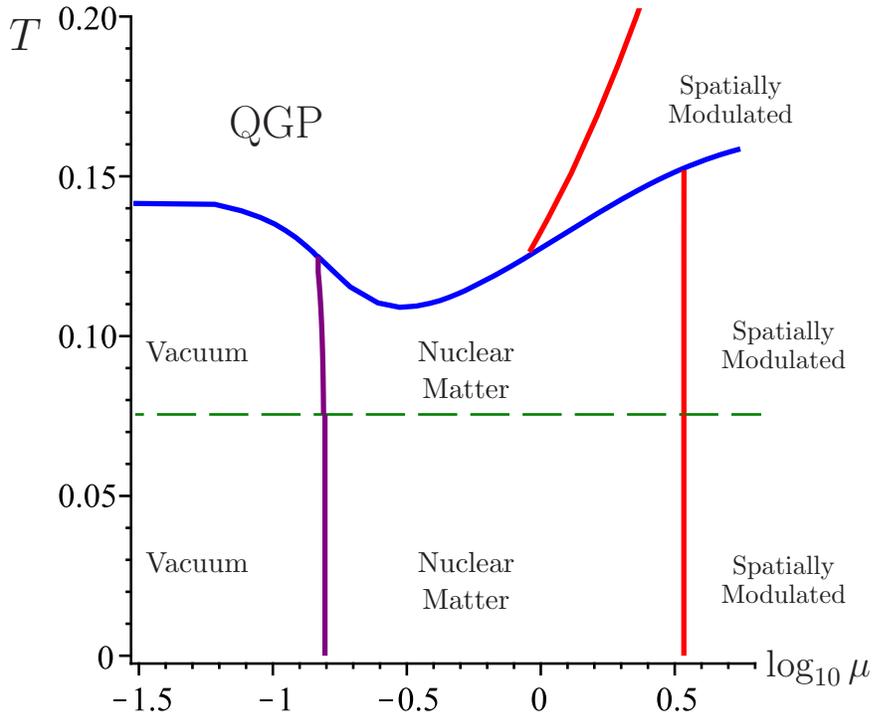, width=10cm}
  \put(-285,265){\makebox(0,0){\LARGE $T$}} 
  \put(15,25){\makebox(0,0){\Large $\log_{10} \mu$}} 
  \put(-220,145){\makebox(0,0){Vacuum}} 
  \put(-120,145){\makebox(0,0){ $\mathrm{Nuclear}$}} 
  \put(-120,131){\makebox(0,0){ $\mathrm{Matter}$}} 
  \put(0,152){\makebox(0,0){ {\small Spatially}}} 
  \put(0,142){\makebox(0,0){ {\small Modulated}}} 
  \put(-190,230){\makebox(0,0){\LARGE $\mathrm{QGP}$}} 
  \put(-20,245){\makebox(0,0){ {\small Spatially}}} 
  \put(-20,235){\makebox(0,0){ {\small Modulated}}} 
  \put(-220,65){\makebox(0,0){Vacuum}} 
  \put(-120,65){\makebox(0,0){ $\mathrm{Nuclear}$}} 
  \put(-120,51){\makebox(0,0){ $\mathrm{Matter}$}}
  \put(0,63){\makebox(0,0){ {\small Spatially}}} 
  \put(0,53){\makebox(0,0){ {\small  Modulated}}} 
\caption{Phase diagram of  our holographic QCD model at finite
  temperature and finite baryon density.
The horizontal dashed line around $T=0.075$
separates the low temperature (LT) and the high
temperature (HT) phases.
The left two purple lines, almost vertically straight
in both of LT and HT phases,
separate the nuclear matter phase and the vacuum phase
where the baryon density vanishes.
The horizontal curved blue line in the high temperature phase,
starting from $T \simeq 0.14$,
is the phase
boundary between the nuclear matter or the vacuum phase,
and the QGP phase.
This line is determined thermodynamically.
The three red lines on the right,
inside the nuclear matter phase of both LT and HT phases
and also in the QGP phase,
denote the onset of
dynamical instability in each phase.
On the right of these lines, the system may be in a spatially
modulated phase, though we do not discuss the stability of this
modulated phase in the paper.
The parameters (whose meaning will be clear in Section \ref{sec:model}
 and \ref{sec:numerical}) are set
to be  $R=1$, $L=0.53$ and $\kappa \simeq 1.018$.}
\label{fig:phase1}
\end{figure}

On the other hand, the study of the gravity duals of large $N_c$
${\cal N}=4$ Super-Yang-Mills(SYM) or QCD are shown to be able to
capture many essential features of real QCD at high temperature
and finite density such as what happened at RHIC. Indeed, a pioneering work by
Domokos and Harvey\cite{Domokos:2007kt} has shown that the bulk Chern-Simons term
will induce a dynamical instability to form the DGR-like modulated phase in the zero
temperature holographic QCD from a bottom-up approach. Recently the
interest in the dynamical instability induced by the Chern-Simons term
is revived by Nakamura, Ooguri and Park in \cite{OO}. They have
explicitly shown that the new modulated phase is the helical wave of
a holographic $U(1)$ current. These motivate us to study the same
issue in the top-down approach of the holographic QCD, i.e.,
Sakai-Sugimoto model\cite{SS1,SS2} at finite
temperature\cite{Aharony:2006da} and finite density\cite{Bergman}. The
advantage of the top-down approach enables us to understand how
marginal the above dynamical instability is in a more consistent and
systematical way. 
In our model we figure out a window for
modulated phase of chiral helical wave at high temperature case as
shown in Fig. \ref{fig:phase1}.

In the next section, we will briefly review the holographic QCD proposed
by Sakai and Sugimoto\cite{SS1,SS2}, and then study thermodynamics of
its generalization to the case with finite temperature and finite
baryon density\cite{Bergman}. 
In Sakai-Sugimoto model the 
meson is dual to a pair of connected $D8$ and $\overline{D8}$ branes,
and the baryon is dual to $D4$ branes wrapping on the internal
$S^4$\cite{WittenBaryon}.
 There is a pulling force on $D8$ exerted by
the wrapped $D4$ whose strength is
proportional to the baryon density. This geometrizes the change of
QCD vacuum by the presence of finite baryon density.  In section
3 we discuss the dynamical instability in holographic QCD and derive
the master equation for it, and in
section 4 we discuss our numerical analysis to find out such an
instability. The physical result of our analysis is summarized in the
phase diagram as shown in Fig. \ref{fig:phase1}.
Finally, we conclude the paper in section 5.  


\section{The Model of Holographic QCD}
\label{sec:model}

The Sakai-Sugimoto(SS) model consists of $N_f$ pairs of $D8$ and
$\overline{D8}$ branes embedded in the near horizon geometry of $N_c$
$D4$ branes on a Scherk-Schwarz circle, which completely breaks the
supersymmetry.
The size of the circle basically controls the mass scale of the
unwanted fields. The configuration can be summarized as follows:
\begin{eqnarray}
\begin{array}{ccccccccccc}
& 0 & 1 & 2 & 3 & (4) & 5 & 6 & 7 & 8 & 9 \\
N_c \ D4 & \times & \times & \times & \times & \times &&&&& \\
N_f \ D8\overline{D8} & \times & \times & \times & \times &  & \times & \times & \times &\times & \times \\
\end{array}
\label{D4D8}
\end{eqnarray}
The $x^4$ direction is the Scherk-Schwarz circle with the anti-periodic
boundary condition for fermionic fields.

The probe $D8$ branes have the embedding function $x_4=x_4(U)$ and
extend in the other 8 dimensions, where $U$ is the radial coordinate of
the background $D4$ brane geometry.
The embedding function can be determined by solving the equations of
motion of the $D8$ DBI action. In \cite{SS1,SS2} the authors found
that there is a smooth interpolation of the $D8$-$\overline{D8}$ pair
from the DBI action analysis, which was then interpreted as the chiral
symmetry breaking of QCD.
In this phase, the background geometry is described by
(\ref{eq:D4bg}), the dual effective QCD is
confined\cite{Aharony:2006da} and the degrees of freedom are hadronic.
By changing the time direction to a thermal circle, we can
introduce temperature into the model.
At a critical temperature, the geometry has a phase transition and
becomes another geometry with horizon\cite{Aharony:2006da,Witten:1998zw}.
This is the high temperature phase that corresponds to the deconfined
phase of QCD and will be described in section \ref{sec:highT} in detail.

In $AdS_5/CFT_4$ correspondence,
the baryons are the $D5$ branes wrapping on the $S^5$.
The RR-flux is coupled to the gauge field on the $D5$
worldvolume and becomes a source for the gauge field.
In order to cancel the tadpole, there will be $N_c$ strings attached
to the $D5$ which looks like a pointlike particle in
$AdS_5$\cite{WittenBaryon}.
Similar constructions
of holographic baryons were done for SS model, which is carried out by
wrapping $D4$ branes on the internal
$S^4$\cite{Hata:2007mb,Hong:2007kx}, and further smearing the $D4$
charges in the non-compact directions for the discussions of QCD phase
diagram\cite{Bergman}.
The $D4$ charge density then plays the role of baryon number density
in the dual QCD.
From the point of view of the probe $D8$, the
wrapped $D4$'s are also regarded as instantons on the $D8$
worldvolume.
However, it turns out that the complete treatment of the baryons as
the world-volume instantons is quite difficult and is yet unknown.
Thus we include the energy contribution of the instanton
as D4-brane DBI action, which was done in  \cite{Bergman}.
Below we will review this results
that will provide the background configuration for our further
study of modulated phase instability.

After wrapping the $D4$, we also have to include the Chern-Simons(CS) action on the $D8$ brane
\be\label{D8CS}
S_{CS} = T_8 \int C_{p+1} \wedge \text{tr} \ \text{exp}(2 \pi \alpha' F)
\ee
if there is non-trivial RR potential $C_{p+1}$ generated by Dp-brane
or non-trivial gauge curvature $F$ on the
$D8$ worldvolume. 
We will regard the wrapped $D4$ branes as sources to
generate non-trivial $F$ along $1,2,3,U$ directions and
focus on the coupling of $U(1)$ part to $SU(N_f)$ part of the gauge
field on $D8$ brane.  The $SU(N_f)$ part
of the gauge field is the anti-self dual connection
whose second Chern character is the number of $D4$ branes, i.e.,
$\frac{1}{8 \pi^2} \text{tr} \ F^2 = N_f n_4 \delta(U-U_c)
d^3x dU$, where $n_4$ is the dimensionless instanton number density
and we also take the total number of the instanton to be a multiple of $N_f$.

\subsection{Thermodynamics of the low temperature phase}

We first consider the low temperature (LT) phase of QCD in which the
quarks and gluons are confined, and the chiral symmetry is
spontaneously broken. The $D4$ background solution dual to the low
temperature phase of QCD is given by \cite{Witten:1998zw},
\begin{eqnarray}
\label{eq:D4bg}
&&ds^2=\Big(\frac{U}{R}\Big)^{\frac32}(\eta_{\mu\nu}dx^{\mu}dx^{\nu}+h(U)dx_4^2)
+\Big(\frac{R}{U}\Big)^{\frac32}\Big(\frac{dU^2}{h(U)}+U^2d\Omega_4^2\Big) \nonumber \\
&&F_4={(2\pi)^3\ell_s^3 N_c \over \Omega_4}\epsilon_4,~e^{\phi}=g_s\Big(\frac{U}{R}\Big)^{\frac34},~h(U)=1-\frac{U^3_{KK}}{U^3},~ x_4 \sim x_4 + \frac{4 \pi R^{3/2}}{3 U_{KK}^{1/2}}, \label{zeroT}
\end{eqnarray}
where $\mu,\nu=0,1,2,3$, $U$ is the radial direction, the $d\Omega_4^2$ and $\epsilon_4$ are the metric and volume form of the $S^4$ of unit radius, and $R^3=\pi g_s N_c \ell_s^3$. This geometry has a cigar shape terminated at $U=U_{KK}$.

The DBI action and CS action in terms of the $D8$ brane embedding
function $x_4(U)$ and the electric field $E(U):=-A_0'$ are the
following \footnote{%
In the following we re-scale the coordinates
  $(U,x_0,x_1,x_2,x_3)$ by factor of $R$ so that they are
  dimensionless, and also rescale $U(1)$ part of the gauge fields by
  factor of ${\sqrt{2N_f} R \over 2\pi \alpha'}$.
Also note that
these actions describe the upper half of the connected D8-branes,
and then the actual action is twice of this.
To investigate the phase structure, it is sufficient to look at this
half and we will consider only this part of the action.}
\bea\label{DBI}
&& S_{DBI} = - N \int dU U^4 \sqrt{h x_4^{'2} +(1/U)^3(1/h -A_0^{'2})}\ , \\ \label{CS}
&& S_{CS} =  {N_c\over 8\pi^2} \int_{M_4\times R_+}A_0 \text{tr}\, F^2 = N n_b \int dU \delta(U-U_c)A_0\ ,
\eea
where the $'$ is the derivative with respect to $U$,  $\beta:=\int
dx_0$ is the inverse temperature, $V_3:= \int dx_1 dx_2 dx_3$,
$\Omega_4$ is the volume of unit $S^4$, $N:={T_8 R^5 N_f V_3 \beta
  \Omega_4\over g_s}$, and the dimensionless baryon number density is
$n_b := {N_c N_f n_4 V_3 \beta \over 2\pi \alpha' R^2 N}$.
Here, for the DBI action, we only include the $U(1)$ part of
the gauge field and neglected the contribution from the $SU(N_f)$
part.\footnote{%
It is not difficult to see that the contributions from this
non-Abelian part is completely mixed, in a complicated way,
with the Abelian part functions we will now solve,
and also the complete form of the contribution is yet unknown.
}
The contribution from the $SU(N_f)$ part is basically the energy of
the instantons and this will be included as D4-brane action on $S^4$ 
when the stationary condition is considered.
The D4-brane action on $S^4$ at $U=U_c$ is given by
\begin{align}
  S_{D4}=& - 
N_4 T_4 \int d\Omega_4 d\tau  e^{-\Phi} \sqrt{-\det g}\bigg|_{U=U_c}
= \frac{N n_b}{3} U_c \,,
\end{align}
where $N_4=N_f n_4 V_3 /R^3$ is the number of D4-branes.

  One can derive the equations of motion from \eq{DBI}, and then solve for $x_4(U)$ and $E(U)$:
\be\label{lowT1}
x'_4=\pm U^{-\frac32}\frac{\sqrt{H_0 \sin^2\theta_c}}{h \sqrt{H-H_0\sin^2\theta_c}}\;,
\qquad E=- n_b U^{\frac32}\frac1{\sqrt{H-H_0\sin^2\theta_c}}\;,
\ee
where
\begin{equation}\label{functionH}
H(U)=U^3 h(U)\; (U^5+n_b^2),~H_0=H(U_c),
\ \text{tan}\ \theta_c = \sqrt{g_{44}/g_{UU}} \ x_4' |_{U=U_c}\ ,
\end{equation}
where $g_{44}$ and $g_{UU}$ refer to the corresponding factors in the
metric (\ref{eq:D4bg}).

\begin{figure}[t]
\begin{center}
\includegraphics[width=12cm]{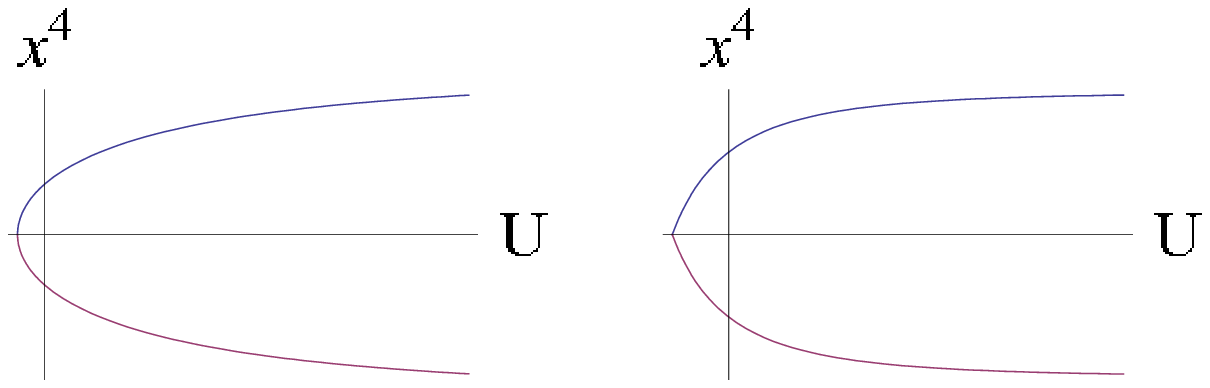}
\caption{\; Smooth $D8-\overline{D8}$ for $n_b=0$ \qquad \qquad  $D8-\overline{D8}$ with a cusp for nonzero $n_b$}
\label{cusp}
\end{center}
\end{figure}

To further solve \eq{lowT1}, we need to impose the fixed length condition, i.e.,
\be\label{lengthfix}
L=\int^{\infty}_{U_c} dU x'_4 ,
\ee
where $2L$ is the UV separation of $D8$ and $\overline{D8}$ along
$x_4$-direction.
The $D8$-$\overline{D8}$ profiles for zero and non-zero $n_b$ are
shown in Fig. \ref{cusp}. Note that this parameter $L$ is particular
to the SS model and does not correspond to any physical
parameter in the real QCD.
We will set $L=0.53$ and $U_{KK}=0.1$ throughout our analysis.
This choice is partly motivated by the physical parameter choice of
SS model and will be explained more in Section \ref{sec:numerical}.

Geometrically, the above solution arises from the balance between the
D8's tension from the DBI part and the pulling by $D4$ wrapping on
$S^4$.
 As a result, the $D8$ profile develops a cusp at IR end $U=U_c$
with a cusp angle $\theta_c$ appearing as a constant of integration in
\eq{lowT1} and \eq{functionH}. This is determined by the force
balance solution
\be\label{forcebalance}
  \Big( \frac{1}{\sqrt{g_{UU}}} \frac{d H_{D8}}{d U_c}+
  \frac{1}{\sqrt{g_{UU}}} {d S_{D4}^E\over d U_c}\Big)_\text{on-shell}=0\;,
\ee
where  $H_{D8} = -S_{DBI} + \int dU \Pi_{A_0} A_0'$,
$S_{D4}^E$ denotes the Euclidean D4 DBI action with
$\Pi_{A_0}$ the conjugate momentum of $A_0$,
and the total
derivative is defined as ${d\over dU_c}:={\partial \over \partial
  U_c}+\Big({\partial x'_4(U) \over \partial  U}\Big|_{\substack{U=U_c\\ n_b, L \; fixed}} \Big) \cdot {\delta
  \over \delta x'_4(U)}$ which should be performed before imposing the
on-shell condition. Solving \eq{forcebalance}\footnote{To solve it,
  one may need to use the trick pointed out in the footnote 6 of
  \cite{Lin}.} for the cusp angle $\theta_c$  yields
\be\label{angle1}
\text{cos}^2 \ \theta_c = \frac{h(U_c)}{9}\frac{n_b^2}{n_b^2 + U_c^5}.
\ee
Note that as $n_b$ reduces to 0, $\theta_c=\pi/2$ and the cusp becomes
a smooth tip.

\begin{figure}[t] 
  \centering
   \includegraphics[width=0.3\textwidth,clip]{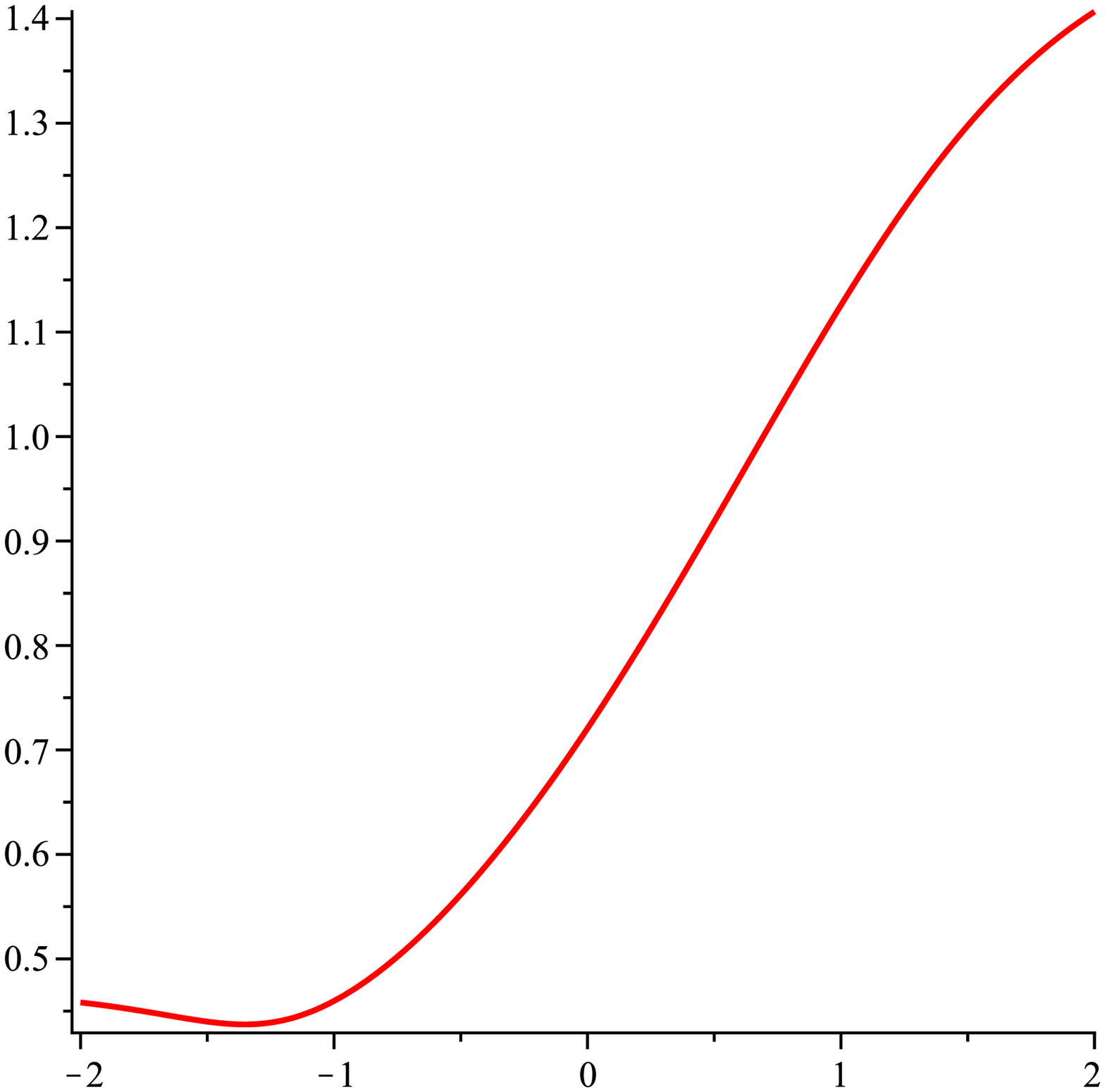}
\hspace{1em}
   \includegraphics[width=0.3\textwidth,clip]{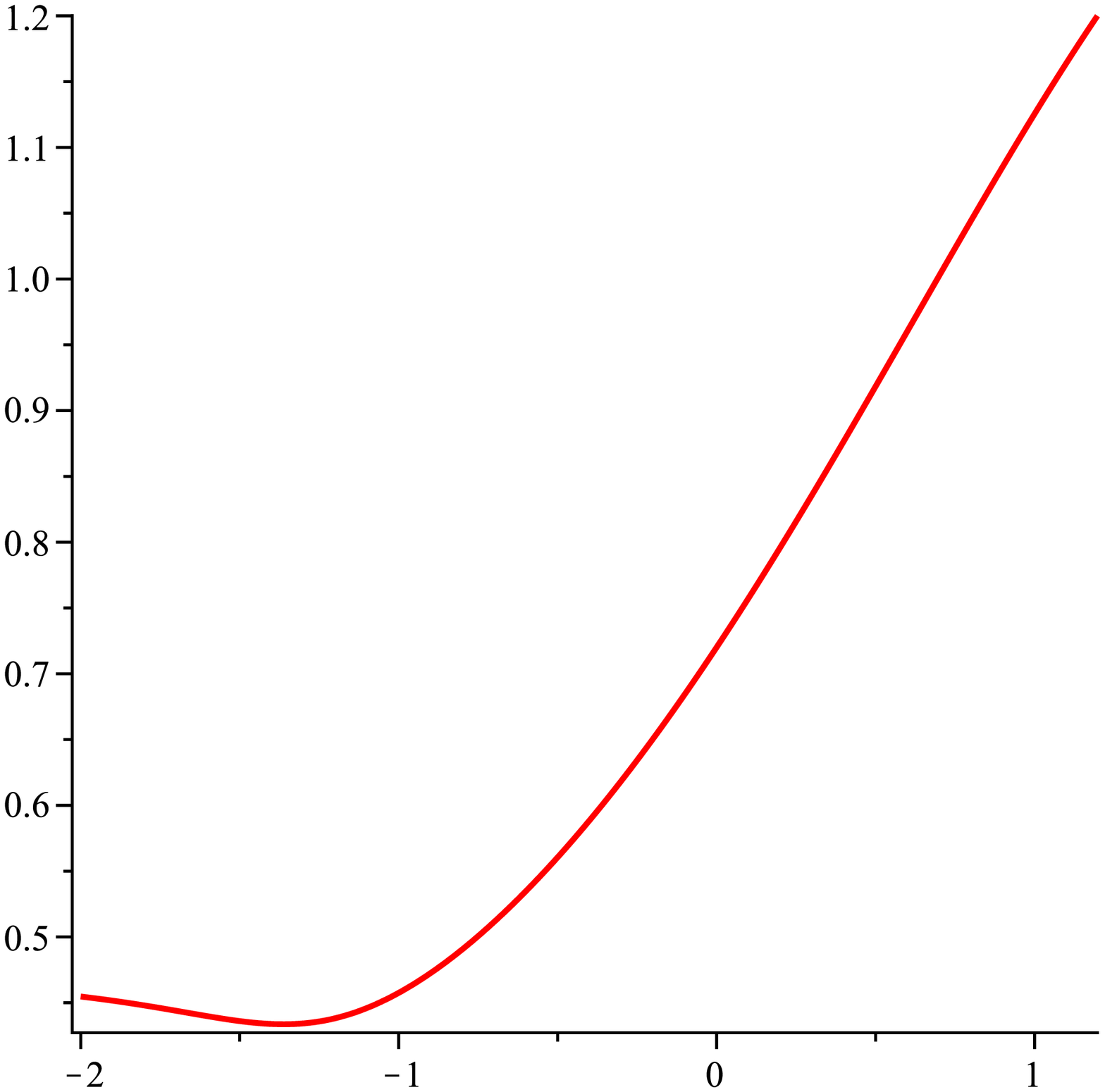}
\hspace{1em}
   \includegraphics[width=0.3\textwidth,clip]{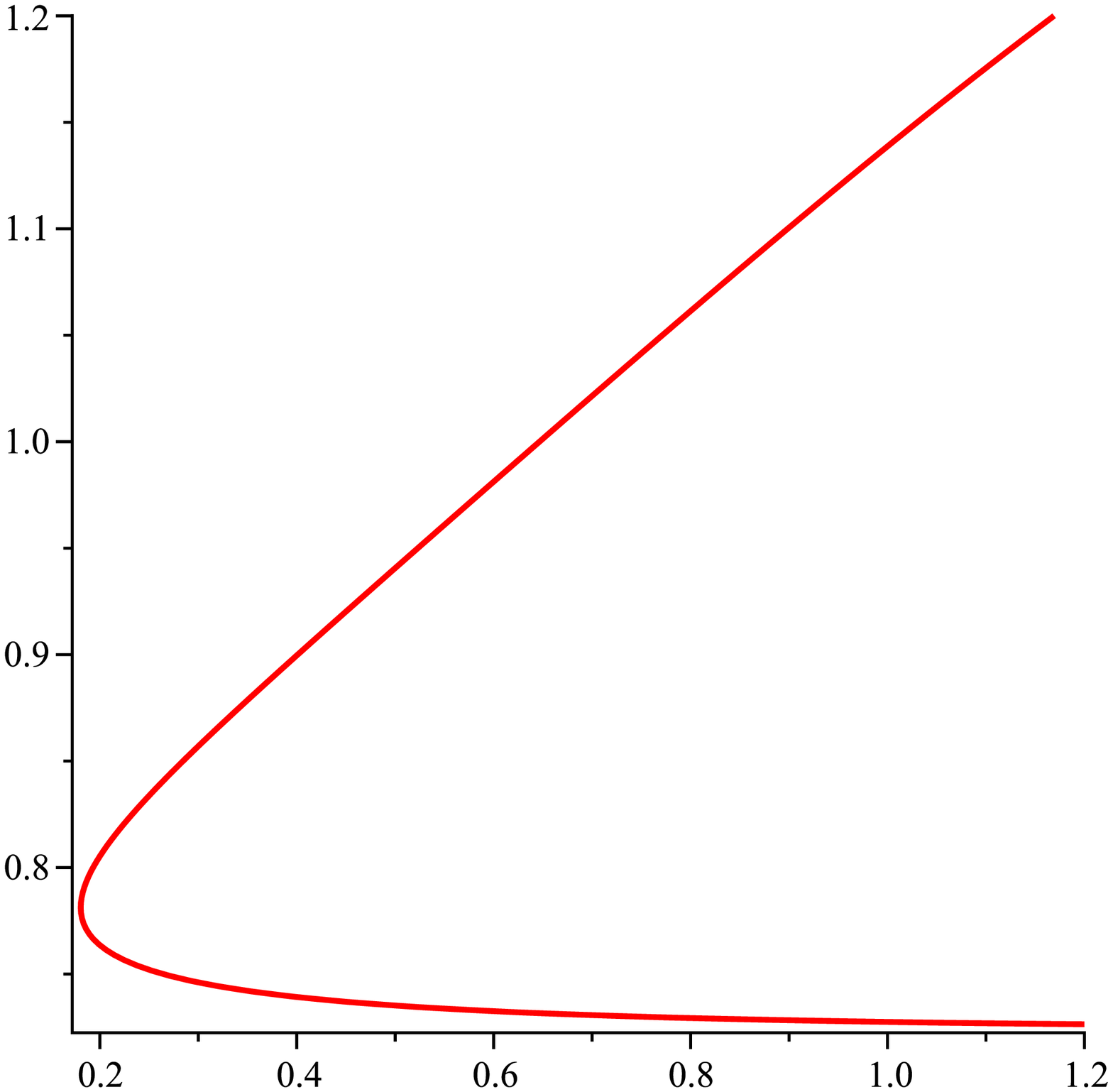}
    \put(-455,150){\makebox(0,0){\small $U_c$}} 
    \put(-320,0){\makebox(0,0){\small $\log_{10} n_b$ }}
    \put(-385,-5){\makebox(0,0){ $(a)$}}
    \put(-285,150){\makebox(0,0){\small $U_c$}} 
    \put(-160,0){\makebox(0,0){\small $\log_{10} n_b$ }}
    \put(-220,-5){\makebox(0,0){ $(b)$}}
    \put(-130,150){\makebox(0,0){\small $U_c$}} 
    \put(0,0){\makebox(0,0){\small $\log_{10} n_b$ }}
    \put(-52,-5){\makebox(0,0){ $(c)$}}
  \caption{The cusp position $U_c$ against $\log_{10} n_b$ at $U_{KK}=0.1$,
    $L=0.53$ for (a) the low temperature phase; and (b)
and (c) are the high
    temperature phase at $T=0.0755$ (or $U_T=0.1$)
and $T=0.1997$ ($U_T=0.7$), respectively. (See sec. 2.2).  }
  \label{Uc_nb_full}
\end{figure}

For a given $n_b$, the cusp location $U_c$ can be determined by solving the
fixed length condition (\ref{lengthfix}).
This relation is displayed in Fig. \ref{Uc_nb_full}(a).
As $n_b$ increases, D4-brane pulls the cusp towards the tip of the
cigar-like space, but at some point the tension of D8-brane overcomes
the pulling force and the position of the cusp goes to large $U$ region.

Next, we discuss the thermodynamics of the $D8$-$\overline{D8}$
branes. We will neglect the back reaction from the probe branes so
that the contribution to the free energy from the $D4$ background
geometry will be the same for different configurations of probe
branes, and can be omitted and considered separately.
Therefore, the  grand canonical potential
of holographic QCD is given by \cite{Gibbons:1976ue}
\be \label{Omega}
 \Omega_{D8}(T,\mu)=\frac{1}{N} \left(\left. S_{DBI}^E\right|_{\mathrm{on-shell}} +\left.  S_{CS}^E\right|_{\mathrm{on-shell}}\right) ,
\ee
where $E$ denotes the Euclidean action,
while the Helmholtz free energy is\footnote{
In the second line of (\ref{HF}), $A_0'$ had been replaced by $x_4'$
by means of the equation of motion. Here we choose to express
$F(T,n_b)$ by $x_4'$ for the convenience of calculating the chemical
potential $\mu$. In the following, (\ref{muthermal}) is carried out as
a chain-rule differentiation since $F$ here is a on-shell
quantity. Equivalently, one can also treat $F$ as off-shell and use
the functional variation method to derive $\mu$. The result is the
same.}
\bea \label{HF}
 F_{D8}(T, n_b) &=& \mu n_b + \Omega_{D8}(T,\mu) = \frac{1}{N}\left. S_{DBI}^E\right|_{\mathrm{on-shell}} + n_b \int_{U_c}^{\infty} A_0'\, dU  \nonumber \\
  &=& \left.\int_{U_c}^{\infty} U^4
\sqrt{\left(h(U) (x_4')^2 + \frac{1}{U^{3}h(U)} \right) \left(1+\frac{n_b^2}{U^5}\right)} dU
 \right|_{x_4'\, \mathrm{on-shell}} \, ,
\eea
where 
$\mu$ is the chemical potential defined by
\be \label{mu}
 \mu := A_0 (U\to \infty) \, .
\ee
Note that the temperature dependence in the LT phase is trivial
and the following expressions have actually no $T$ dependence in this phase.
The system contains also wrapped D4-branes, and its free energy is
simply given by the Euclidean action (normalized by the
factor $N$).
Thus the total free energy is
\begin{align}
\label{total_F}
  F(T, n_b) =& F_{D8}(T, n_b) + \frac{1}{N} S_{D4}^E \,.
\end{align}

From thermodynamics, the chemical potential is also obtained by
\begin{align}
  \label{muthermal}
 \mu =& \left. \frac{\partial F}{\partial n_b} \right|_{T,L} 
\nn\\=& 
\left.\frac{\partial F_{D8}}{\partial n_b}\right|_{x_4', U_c, T, L}
+ \left.\frac{\partial F_{D8}}{\partial x_4'} \frac{\partial x_4'}{\partial
    n_b} \right|_{U_c,T,L} 
+ \left.
\left( \frac{\partial F_{D8}}{\partial U_c}
      + \frac{1}{N}\frac{\partial S_{D4}^E}{\partial U_c}
\right)
 \frac{\partial U_c}{\partial n_b}\right|_{x_4',T,L}
+\left.\frac{1}{N} \frac{\partial S_{D4}^E}{\partial n_b} \right|_{x_4',T,L}
\,.
\end{align}
Note that this formulation is under the implicit constraint of $L$
being fixed. The second term on the right hand side vanishes, because
$\frac{\partial F}{\partial x_4'}$ gives rise to an constant of motion
inside the integral, and $\left. \int_{U_c}^{\infty} \frac{\partial
    x_4'}{\partial n_b} \right|_{U_c} = \frac{d L}{d n_b} = 0$ since
$L$ is fixed\cite{Bergman}.
The third term also vanishes due to the force balance condition
(\ref{forcebalance}) (note that $N F_{D8}=H_{D8}$).
 The remaining terms are consistent with the fixed length condition.
 As a result, the chemical potential reads
\footnote{c.f. footnote 8 of \cite{Lin}.}
\be \label{muformula}
 \mu = \int_{U_c}^{\infty} A_0' dU +
\frac{1}{N} \frac{\partial S_{D4}^E}{\partial n_b}
=\int_{U_c}^\infty dU \,
n_b \sqrt{\frac{U^3}{H(U)-H_c \sin^2 \theta_c}} + \frac{U_c}{3}
 \,.
\ee

\begin{figure}[htb]
  \centering
\includegraphics[width=0.3\textwidth,clip]{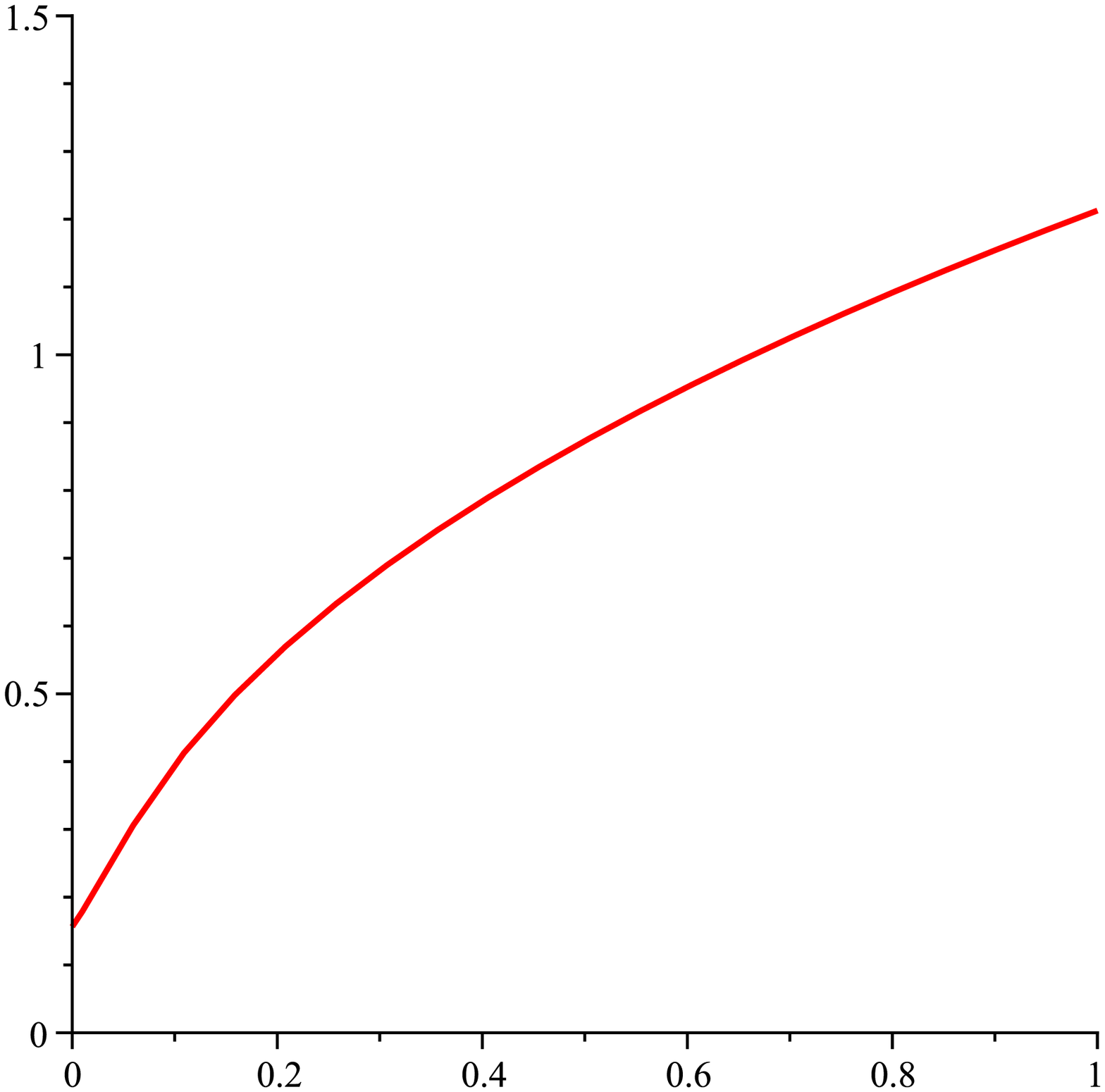}
 \hspace{1em}
\includegraphics[width=0.3\textwidth,clip]{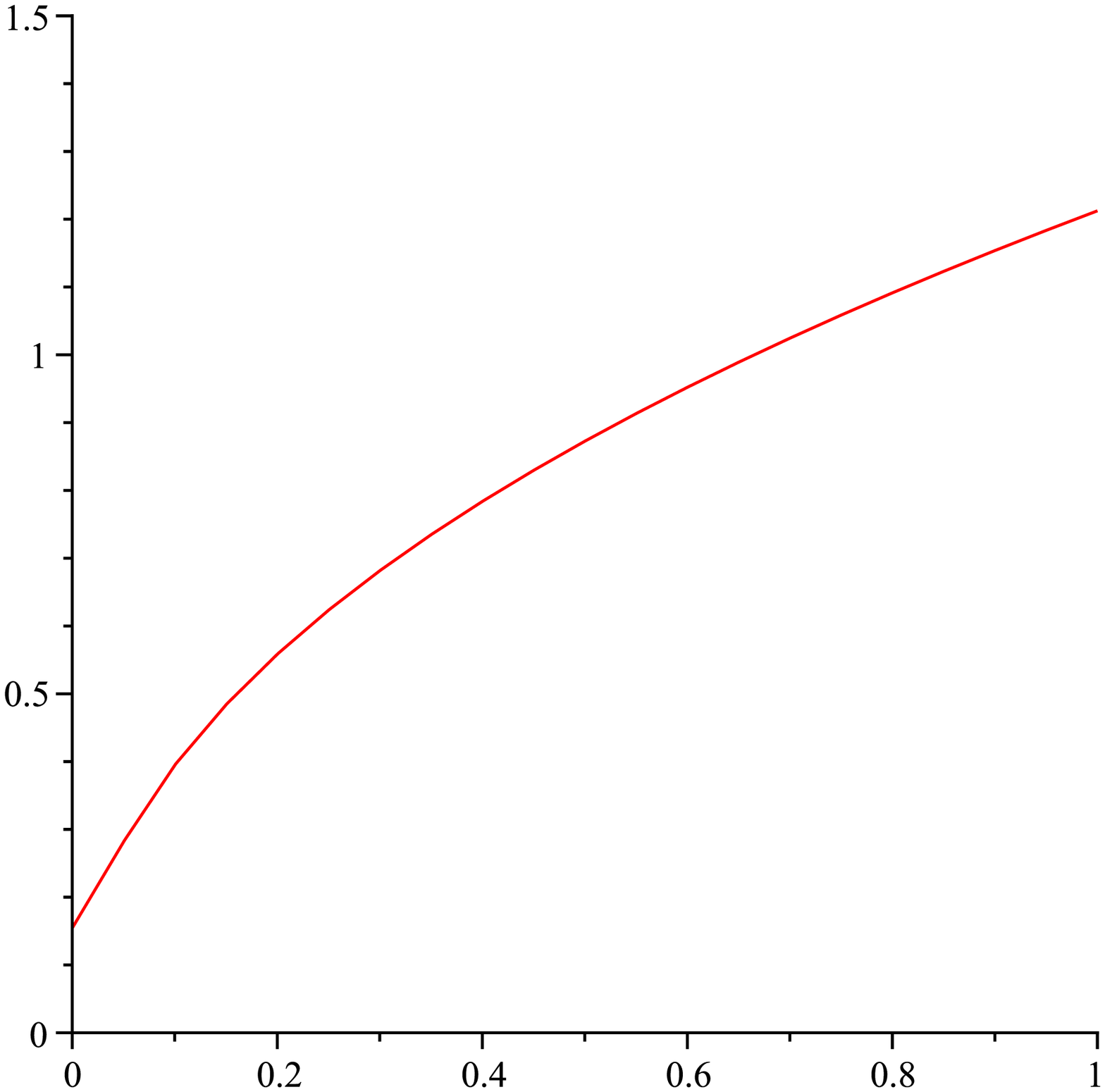}
 \hspace{1em}
\includegraphics[width=0.3\textwidth,clip]{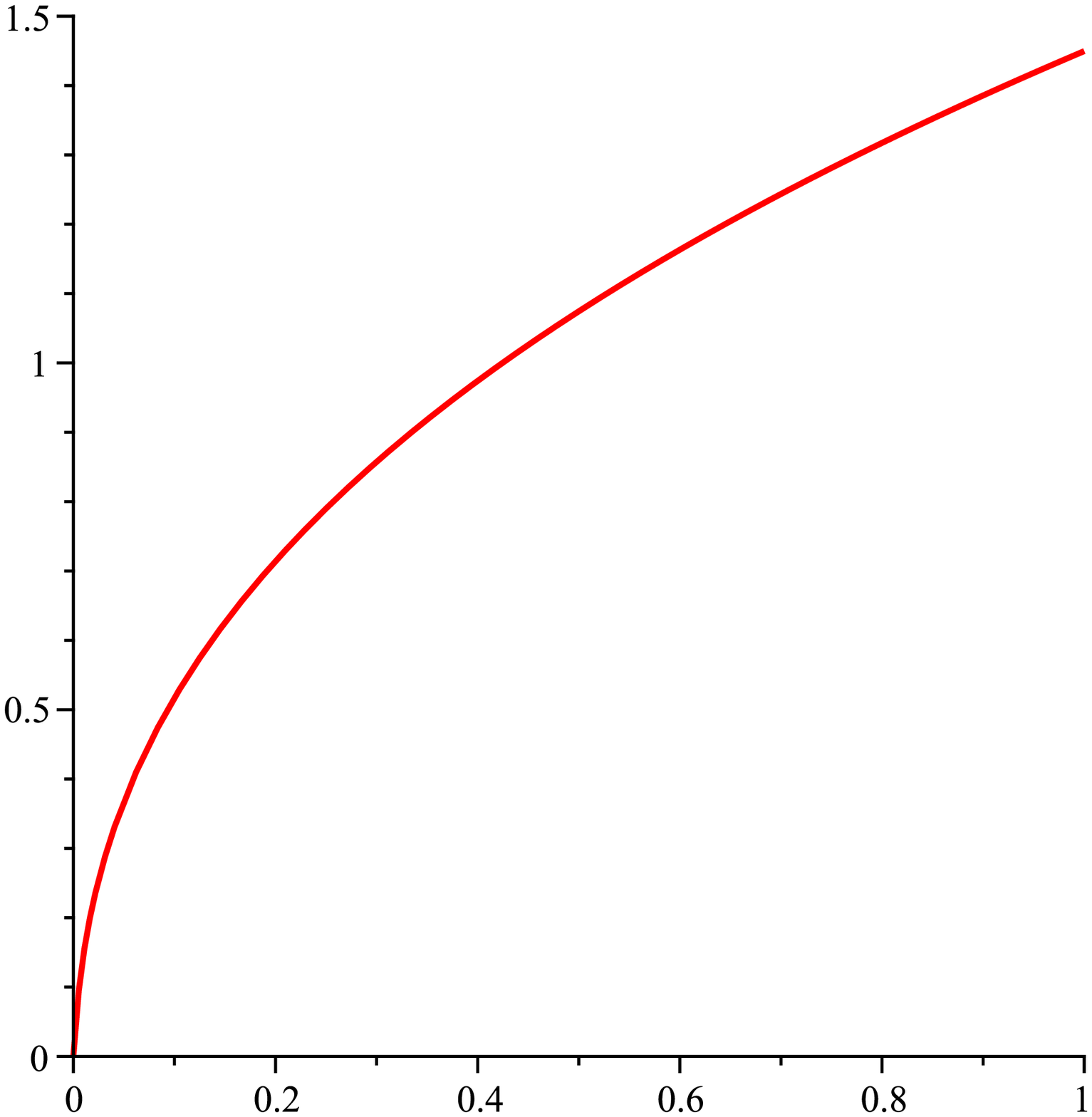}
     \put(-465,135){\makebox(0,0){ $\mu$}}
     \put(-317,-5){\makebox(0,0){ $n_b$}}
     \put(-385,-13){\makebox(0,0){ $(a)$}}
     \put(-308,135){\makebox(0,0){ $\mu$}}
     \put(-160,-5){\makebox(0,0){ $n_b$}}
     \put(-225,-13){\makebox(0,0){ $(b)$}}
     \put(-143,135){\makebox(0,0){ $\mu$}}
     \put(0,-5){\makebox(0,0){ $n_b$}}
     \put(-65,-13){\makebox(0,0){ $(c)$}}
  \caption{The chemical potentials versus $n_b$ in: (a)the low
    temperature phase, (b) the high temperature nuclear
    matter phase at $T=0.0755$(See sec. 2.2)
and (c) the QGP phase at $T=0.0755$(See sec. 2.2).
In all the figures, the horizontal line denotes $n_b$ and the vertical
one $\mu$.}
  \label{mu_all}
\end{figure}

 The chemical potential $\mu$ as a function of $n_b$ is displayed
 in Fig. \ref{mu_all}(a).
As $n_b$ goes to zero, $\mu$ approaches a nonzero value.
Below this value, there is no cusp configuration and then the system
is described by the $D8$-$\overline{D8}$ configuration without
instantons.
This is called the vacuum phase\cite{Bergman}.
In this phase, $\mu$ is a free parameter and we compare the grand
canonical potential of the nuclear matter phase and the vacuum phase,
and conclude that as long as the cusp configurations exist, the
nuclear matter phase is always favored.

\subsection{Thermodynamics of the high temperature phase}
\label{sec:highT}
   Similarly, one can also consider the high temperature (HT) phase of holographic QCD, which is dual to the following metric\cite{Aharony:2006da}
\bea\label{HTbg}
ds^2&=&
\left( {U\over R} \right)^{3/2}[-h(U)dt^2+dx_1^2+dx_2^2+dx_3^2+dx_4^2]
+\left( {R\over U} \right)^{3/2}
\left[ {dU^2\over h(U)}+U^2d\Omega_4^2 \right] , \nonumber \\
h(U)&=&1-{U_T^3\over U^3} .
\eea
The dilaton and the 4-form flux are the same as in \eq{zeroT}. The
Hawking temperature determined from the metric is $T:={3\over
  4\pi}\sqrt{U_T\over R^3}$ which is the temperature for holographic
QCD.

The transition between the LT and the HT phases occurs at $T_c
={3\over 4\pi}\sqrt{U_{KK}\over R^3} \simeq 0.0755$
\cite{Aharony:2006da}.
 Above $T_c$ the temperature is varied by tuning $U_T$, and the dual
 QCD is deconfined. This critical temperature is obtained by comparing
 the free energies of the supergravity background of each phase. This
 is the background geometric transition which happens in the leading
 order at large $N_c$. Since the effect from the $N_f$ quarks can be
 neglected in the planar limit, the boundary between the HT and LT
 phases is straight along the axis of $\mu$, as shown by the dashed
 line in Fig.~\ref{fig:phase1}.

Embedding the $D8$ and $\overline{D8}$ branes in (\ref{HTbg}), by the
same procedure in the LT case we can solve the $D8$-brane profile and
the electric field, now together with the D4-brane action
$S_{D4}^E = \frac{N n_b}{3} \sqrt{h(U_c)} U_c$.
The cusped solution is given by
\be\label{highT1}
x_4'=\pm \; U^{-3/2} \sqrt{H_0 \sin^2\theta_c \over h (H-H_0\sin^2\theta_c)}\;,
\quad
E=-n_b U^{3/2} \sqrt{h \over (H-H_0\sin^2\theta_c)}
\quad
\mathrm{(Nuclear ~ matter)} \; .
\ee
The function $H$ and the angle $\theta_c$ are also given by
\eq{functionH} and \eq{angle1}, with $h$ taken from
(\ref{HTbg}). Since the $D8$ and $\overline{D8}$ are joined at $U_c$
with a cusp angle $\theta_c$, with broken chiral symmetry this
solution describes the nuclear matter in the dual QCD. Variation of
$U_c$ with $n_b$ at $T=0.0755$ (i.e., $U_T=0.1$)
and $T=0.1997$ ($U_T=0.7$) are displayed in
Fig.~\ref{Uc_nb_full}(b) and (c).
For lower temperature, the behavior is similar to that of the low
temperature phase, and there always exists a solution.
For higher temperatures, there will be no solution of the fixed length
condition for small $n_b$, but these temperatures turn out to be in
the QGP
phase, which is described below, and then we will not face this
missing solution problem.

The grand canonical potential, the Helmholtz free energy, and the
chemical potential in the HT nuclear matter phase can be obtained
formally from (\ref{Omega}), the first line of (\ref{HF}) and
(\ref{total_F}), and
(\ref{muthermal}) respectively.
The chemical potential is given by
$\mu = \int_{U_c}^{\infty} A_0' dU + \frac{1}{N}\frac{\partial
  S_{D4}^E}{\partial n_b}
=\int_{U_c}^\infty dU \, n_b \sqrt{\frac{U^3 h(U)}{H(U)-H_0 \sin^2
    \theta_c}} + \frac{U_c \sqrt{h(U_c)}}{3}$.
The behavior $\mu$ against $n_b$ is displayed in
Fig.~\ref{mu_all}(b).
Again, $\mu$ goes to a finite value as $n_b$ vanishes,
and there is also the vacuum phase in the high temperature
region.

Moreover, unlike the LT case where the
parallel $D8$-$\overline{D8}$ configuration is not allowed unless
$L=0$, in the HT phase we may have either the the chiral symmetry
broken configuration (\ref{highT1}), or the chiral symmetry restored
one described by the parallel $D8$ and $\overline{D8}$:
\be\label{highT2}
 x_4'=0\;, \qquad
 E=-A_0'= \frac{n_b}{\sqrt{U^5+n_b^2}}
 \qquad
 \mathrm{(QGP)}\;.
\ee
Both branes extend in the radial direction from the horizon $U_T$ to
the boundary. Since the chiral symmetry is restored and the effective
theory is deconfined, the configuration (\ref{highT2}) describes the
QGP phase.
The chemical potential of the QGP phase is 
\be \label{muQGP}
 \mu := \int_{U_T}^{\infty} A_0' dU = A_0 (U \to \infty)\; , 
\ee
where $A_0(U_T)$ is set to 0 on the horizon to satisfy the regularity
requirement.
 In the QGP phase there is of
course no baryon degrees of freedom, so here we should regard the
chemical potential as that of the liberated quarks and anti-quarks
from the dissociation of baryons and mesons. 
Th chemical potential as a function of $n_b$ is displayed in
Fig.~\ref{mu_all}(c).
For this QGP case, $\mu$ vanishes as $n_b$ goes to zero.

To obtain the boundary between the nuclear matter and the QGP phase,
we compare the grand canonical potential $\Omega$,
 which are
$\Omega_{\chi b}(T,\mu)= \int_{U_c}^{\infty}\! dU
U^5 \sqrt{\frac{U^3h(U)}{H(U)-H_0 \sin^2\theta_c}}$ for 
the nuclear matter phase and
 $\Omega_{QGP}(T,\mu)=\int_{U_T}^{\infty} dU
 \frac{U^5}{\sqrt{U^5+n_b^2}}$ for the QGP phase.
Note that the each $n_b$ in the above formulae takes different values
in each phase, for the same $T$ and $\mu$.
From our numerical calculation, we
determine the phase boundary where the difference of the potential
goes to zero.
These are the boundary between
the QGP and the nuclear matter in the high temperature phase in
Figure 1.
In this high temperature phase, there is also the vacuum phase where
the D8-brane does not develop the cusp.
The boundary between the vacuum and the nuclear matter phase is
determined by $\lim_{n_b\rightarrow 0}\mu(n_b)$ for a given
temperature, and by comparing the grand canonical potential the
nuclear matter phase is again preferred if exists.
The boundary between the QGP and the vacuum phase
is also determined by comparing the grand canonical potential.
The potential for the vacuum phase is given by
$\Omega_\text{vac}(T) = \int_{U_c}^\infty dU \, \sqrt{\frac{U^{13}
h(U)}{U^8 h(U) - U_c^8 h(U_c)}}$.

\section{Dynamical Instability}
\label{sec:dynamical_instability}
To see if there is any dynamical instability of holographic QCD at
finite baryon density, we need to consider the bulk fluctuations around
the above background. We then solve the bulk fluctuations and derive
the expectation values of the corresponding operators.
If the expectation value grows with time, it suggests the vacuum is
unstable and implies dynamical instability.

For our purpose, we turn on the fluctuations $y(x_{\mu},U)$ for $D8$
profile and $f_{IJ}(x_{\mu},U):=\partial_I a_J-\partial_J a_I$ for
$D8$ gauge fields. Holographically, the fields $a_i$, $a_U$ and $y$
are dual to the chiral current $\langle J^i_{L,R} \rangle$, pion
field $\langle \bar{q}\gamma^5 q\rangle$ and chiral symmetry
violation $\langle \partial_{\mu} J^{\mu}_{L,R}
\rangle$\cite{Antonyan:2006vw}, respectively. If there are
normalizable unstable (i.e., its magnitude grows with time) solutions
for these bulk fluctuations, it implies the dynamical instability to
new phases. For example, if exists, the modulated phase for $a_i$,
as discussed in \cite{OO}, is dual to helical structure of large $N_c$
QCD; and the new modulated phases for $a_U$ and $y$ will be dual to
the chiral density wave, a kind of DGR instability.

To obtain the field equations for the above bulk fluctuations, we
expand the Lagrangian $L_{DBI}+L_{CS}$ up to quadratic order
for both low  and high temperature phases, and perform the
corresponding variations.
The results are
\be
{U^5 \over L_0}\left\{{\Delta_p \over h} \partial_0^2 y - (1- \Delta_p E^2) \partial_i^2 y - \Delta_p x_4' E \partial_i f_{0i}\right\}-\left\{{{U^8 h}\over L_0}\left(y'-{x_4'\over L_0} L_1  \right)\right\}' =0\;,
\ee
\be
{U^5 \over L_0}\left\{-\Delta_p x_4' E \: \partial_i^2 y + 
\left(\Delta_p (x_4')^2 +{1 \over {U^3 h}}\right) \partial_i f_{0i} \right\}
+\left\{ {U^5 \over L_0} \left( f_{0U} - {E \over L_0} L_1 \right)
\right\}' = 0\;,
\ee
\be
{U^5 \over L_0}\left\{ \partial_0 f_{0U} -{h \over \Delta_p} \partial_i f_{iU} - {E \over L_0}  \partial_0  L_1 \right\}=0 \;,
\ee
\bea
{U^5 \over L_0} \left\{ -\Delta_p x_4' E \: \partial_0 \partial_i y 
+\left( \Delta_p (x_4')^2 + {1 \over {U^3 h}} \right) \partial_0 f_{0i} \right\}
+ {L_0 \over U^3} \partial_jf_{ij} + \left( {{U^5 h} \over {L_0 \Delta_p}} f_{iU} \right)' && \nn \\
-\kappa \epsilon_{ijk}(2 A_0 \partial_j f_{kU} - E f_{jk} + A_0 \partial_U f_{jk})=0&& \hspace{-0.5cm},\label{aieom}
\eea
where $L_0$ and $L_1$ represent the expressions
\begin{align}
L_0 =& U^4 \sqrt{h (x_4')^2 + U^{-3}\left( {1 \over \Delta_p} -
    E^2\right)} \,,
\qquad
L_1 = {U^8 \over L_0}\: (y' x_4'\, h - U^{-3} E\, f_{0U}) .
\end{align}
and the index $p$ labels different phases for the factor $\Delta_p$,
such that $\Delta_{LT}:=h$ and $\Delta_{HT}:=1$.
Note that the formulation of $h$ in the LT and HT
cases should be correspondingly taken from (\ref{eq:D4bg}) and
(\ref{HTbg}) respectively.

Though these are coupled equations which cannot be decoupled in
general, we can isolate the equation for $f_i:={1\over
2}\epsilon_{ijk}f_{jk}$ from the other bulk fluctuations by
applying $\epsilon_{ijk}\partial_j$ to \eq{aieom}.
We then arrive at the master equation
\be
\label{master1}
\left({U^5 h \over {L_0 \Delta_p}} f_k'\right)' + {L_0 \over U^3}  \left\{
- {\Delta_p \over h} \left(1+{E^2 \over L_0^2} U^5\right)\partial_0^2 f_k
+ \partial_i^2 f_k \right\} + 2 \kappa \epsilon_{ijk} E \partial_j f_i=0 \;,
\ee
in which $\Delta_p (x_4')^2 + {1 \over {U^3 h}} = {\Delta_p \over h}
\left({L_0^2 \over U^8} + {E^2 \over U^3}\right)$ is used.
Note that $y$, $a_U$ and $a_0$ are decoupled. This is the similar
equation discussed in \cite{OO} for the instability to form the
modulated phase; however,  the effective CS coupling is not a fixed
constant as in \cite{OO} but given by
\be\label{kappa1}
\kappa:={n_b\over 4\pi^2 n_4}=288\pi^2 {1\over \lambda^2} {U_{KK} \over R}\,,
\ee
and thus it depends on the choice of the parameters like 't Hooft coupling $\lambda:=g_{YM}^2N_c$,
but not on the instanton number density.

The parameter $\kappa$ and the baryon number density $n_b$
will turn out to
govern the strength of instability to form the modulated
phase as discussed in \cite{OO}.
So when $n_b$ and $\kappa$ take a sufficiently large value, the dynamical
instability would appear, as we will confirm by numerical calculation
in the next section.

\section{Numerical Results}
\label{sec:numerical}

In this section, we present our numerical analysis for possible dynamical
instability,  and the physical  result is
summarized in the phase diagram shown in Figure 1.

 We start by summarizing our choice of the parameters.
We set $M_{KK}=\frac{3}{2}\sqrt{\frac{U_{KK}}{R^3}}$ and 't Hooft
coupling $\lambda =g_{YM}^2 N_c$ as
$M_{KK}\simeq 950 \text{MeV}$ and $\lambda \simeq 16.71$
(for example, \cite{Kim:2007zm}).
This choice was set by using $\rho$-meson mass and the pion decay
constant.
We still have a freedom to choose the mass scale by $R$
and then set $U_{KK}/R=0.1$ and all the dimensions are adjusted by including
$R$ appropriately.
This choice leads $\kappa \simeq 1.017980206$,
and we choose the asymptotic half separation $L=0.53$ \footnote{
Note that in this parameter choice, $\lambda$ is not quite large and
there exists a potential danger of having open string tachyons
for too small $L$\cite{SS1}.
Our choice avoids this problem.}.

We numerically solve the
equations of motion derived in the previous section to find unstable modes.
We use the standard "shooting" method to find the normalizable mode to
determine the onset of the instability, which is given by the smallest
available baryon number density
that leads normalizable unstable modes at a given temperature.

First we consider the master equation (\ref{master1}).
We assume the mode expansion
$f_i \!=\! g_i(U) e^{-i(\omega t + k_j x^j)}$.
Then, as explained in \cite{OO},
the differential operator $\epsilon_{ijk}\partial_j$ has the eigenvalues
$\pm k$ and $0$ where $k=\sqrt{k^i k^i}$,
and the equations of motion can therefore
be diagonalized with respect to $i$ independently of $U$.
Note that the mode with zero eigenvalue turns
out to be unphysical since it is in conflict with the Bianchi identity.
Thus we diagonalize $g_i$ and drop the subscript $i$ for $g_i(U)$ from
now on, and these three are distinguished by the eigenvalue that appears in
the equation of motion (\ref{master1}).

We are looking for the normalizable solutions with negative $\omega^2$.
Near the boundary, the asymptotic solution is
\begin{align}
  g(U) \sim m + \frac{\nu}{U^{3/2}} \,,
\end{align}
where the leading constant $m$ part gives the nonnormalizable mode
and  $\nu U^{-3/2}$ term is the normalizable
mode, and $m$ and $\nu$ are constants of integration.
We are going to tune $k$ for given parameters to find a solution that
has vanishing $m$.
Now the connected D8-brane configuration is symmetric under $x_4 
\leftrightarrow -x_4$, and then two independent
modes are either even or odd functions.
This is equivalent to choose
Neumann boundary condition $\{ g(U_c)=1, g'(U_c)=0 \}$ or
Dirichlet one $\{ g(U_c)=0, g'(U_c)=1 \}$ at the connecting
point\cite{SS1}.
Among these choices, we have observed that the first choice is prone
to be unstable for various choices of the parameters.
When we consider the other set of the equations of motion,
we also take the same two boundary conditions for $a_U$ fields,
while for $y$ and $a_0$ fluctuations we need to take the instantons
into account.
Since we have put the wrapped D4-branes at $U=U_c$ and have the
boundary action (\ref{CS}), thus we take the boundary conditions that
do not change this boundary term imposed as the background.
So we take Dirichlet boundary condition for $y$ so that the
position of D4 brane would not change.
Since the conjugate variable of $a_0$ with respect to $U$ is
a constant of motion,
we put Neumann boundary condition for
$a_0$, in order for this constant of motion, which is related
to the instanton number density $n_4$, not to change.
Finally, when we consider the instability for the QGP phase, we
impose the in-falling boundary condition at the horizon, as usual.
 
We first analyze the marginal case $\omega^2=0$, where an unstable
mode would start to appear. 
In the low temperature phase, we tune the baryon number density $n_b$ 
to find the appearance of the instability.
For the high temperature phase, we have another parameter $U_T$
that gives the temperature $T$.

The unstable mode appearing at the smallest $n_b=n_{b(\mathrm{crit})}$ can be
considered to be the onset of the instability. 
The dynamical instability occurs for $n_b > n_{b(\mathrm{crit})}$ at a
given temperature $T$.
Using this criterion, we determine the phase boundary between the
``nuclear matter phase" and the ``modulated phase" in the
$T$--$n_b$ plane.
Note that we have not confirmed that this modulated phase stably
exists, and we focus on drawing the phase boundaries.
In Mathematica we
implement a shooting method to look for the normalizable modes of
$g(U)$ which vanish at large $U$.
The Mathematica code is like the following schematically,
\begin{equation}
{\tt NDSolve[ \{ EOM(k,\omega^2=0) \ of \ g , g[U_c] == 1, g'[U_c] == 0 \}, g, \{ U, U_c, U_{max} \} ] }.
\end{equation}
Note that there are momentum dependences in the equation of motion.
 So we need to set up a do-loop to
run over a range of momentum $k$ and plot the diagram of $k$-$g[U_{max}]$, where
$U_{max}$ is chosen such that the sub-leading terms in $g(U_{max})$ can
be dropped out and only the leading constant term $m$ remains.
 
\begin{figure}[t]
\centering
\epsfig{file=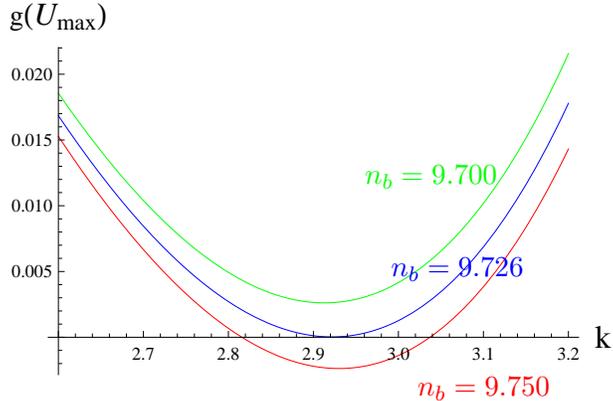, width=8cm}
 \put(-70,75){\makebox(0,0){ \textcolor{green}{$n_b=9.700$}}}
 \put(-60,40){\makebox(0,0){ \textcolor{blue}{$n_b=9.726$}}}
 \put(-50,-5){\makebox(0,0){ \textcolor{red}{$n_b=9.750$}}}
\caption{The numerical cartoon indicates the dynamical instability at
  $\omega=0$ to form the modulated phase in low temperature phase by
  tuning $n_b$; namely, there exists a normalizable solution at
  critical $n_b$. The critical value of $n_b$ is around $9.726$.}
\label{fig:zeroT}
\end{figure}

Here we provide the plot to show how we look for the dynamical
instabilities in the low temperature phase; the plots in the HT phase
at a given temperature are similar.  Fig. \ref{fig:zeroT} shows that
the critical value of $n_b$ at $\omega^2=0$ is around $n_b \approx
9.726$ 
since at this value there is a zero for $g(U_{max})$.
That is to say we can find a normalizable solution at this critical
value with $\omega^2=0$. Repeating this procedure for both low and
high temperature phases, including QGP phase in the high
  temperature phase,
the boundary to
the the modulated phase (due to dynamical instability) is
determined.
Finally we translate these values of $n_b$ into the chemical potential
in the corresponding phase
to the phase boundaries for $\mu$--$T$ diagram.
The result is summarized in Fig. \ref{fig:phase1}.

From our analysis, we see that $n_b$ and $\kappa$ govern the emergence of
the instability, and for a given $\kappa$, which is determined by the
choice of the parameters, we may choose a sufficiently large $n_b$ to
find the onset of the instability.

\begin{figure}[htb]
\centering
\epsfig{file=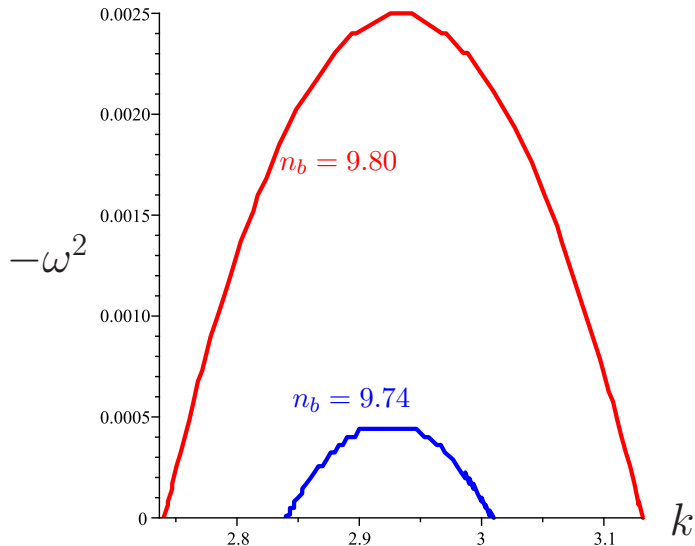, width=7.5cm}
 \put(10,15){\makebox(0,0){\LARGE $k$}}
 \put(-230,115){\makebox(0,0){\LARGE $-\omega^2$}}
 \put(-115,60){\makebox(0,0){\textcolor{blue}{$n_b=9.74$}}}
 \put(-120,150){\makebox(0,0){\textcolor{red}{$n_b=9.80$}}}
\caption{The dispersion of $k$ and $-\omega^2$ for the
  normalizable unstable modes in the low temperature phase. 
 The lower (blue) is for $n_b=9.74$ and
 the higher (red) is for $n_b=9.80$.}
\label{omega-scan-LT}
\end{figure}

  To confirm that the normalizable solutions we obtained indeed indicate
instability, we next check the $k$ dependence of $\omega(k)$.
Fig. \ref{omega-scan-LT} is the plot of $k$ dependence of
$-\omega(k)^2$ at which the normalizable unstable solutions
appear, for $n_b=9.74$ and $n_b=9.80$ in the low temperature phase.
The horizontal axis is $k$ and the vertical axis is $-\omega^2$.
It shows that unstable modes appear between two marginal values of $k$
for $n_b> 9.726$ as seen before.


Finally, we also analyze the instability of the other three modes.
They cannot be decoupled, and we then take the common $\omega=0$ and $k$
for all and search for
the normalizable mode for one of them.
For the instability to appear, all of the fluctuations
have to be normalizable\footnote{
Nonnormalizable modes may become normalizable as the
frequency and the momentum change, however, we do not perform the full
analysis here, but just the preliminary ones.}.
For the QGP phase, there seems no normalizable mode.
In the nuclear matter phases,
we found a normalizable mode with $\omega=0$
according to the combination of
the boundary conditions,
and it could be the onset of an instability.
It appears at rather low baryon densities in general.
However, the position of the onset depends very much on the choice of the
boundary conditions, and we do not have the physical argument to fix
this ambiguity at this moment.
Therefore, we concentrate on the instability corresponding to 
$f_i$ modes we discussed, and do not include into our phase diagram
these might-be onsets of the
instabilities from the other modes.
It will then be nice if one could perform the full analysis and
discuss the instabilities regarding other modes.

\section{Conclusion}

  In this paper we study the dynamical instability of holographic QCD
  at finite density, which is usually
difficult to attack by the conventional perturbative approach or the
first principle lattice simulation due to the sign problem in the
presence of finite chemical potential. One more advantage of
holographic approach is to have the bulk geometric picture to
illustrate the dynamical properties of QCD. For example, both the thermodynamical and dynamical
instabilities are due to the pulling force of the $D4$ baryon branes
exerted on the probe $D8$-$\overline{D8}$ meson branes. Our results are summarized in
the phase diagram Fig. 1. The essential feature in Fig. 1 is the
possible existence of the new modulated phase, which is missed by the
homogeneity condition of thermodynamical consideration.

  A key ingredient for the appearance of such a dynamical instability is that
  the baryon density induces the electric field in the holographic
  $D8$-$\overline{D8}$ brane, which couples to the QCD chiral current
  via the bulk Chern-Simons term.
We find that the emergence of the instability is determined by the
effective Chern-Simons coupling and the value of the baryon number
density.
By tuning the baryon number density high enough, a dynamical
instability would appear, and this is consistent with 
the expectation from the previous
large $N_c$ analysis, at least for sufficient low temperatures.
We have investigated the instabilities for several fluctuations on
$D8$, and
it should also be interesting to take the open string tachyon into
account, which is dual to the chiral condensate
and is an essential feature missed by the original Sakai-Sugimoto model,
and to consider the dynamical instability with it.

Finally, we are left with an open question about the treatment of the
non-Abelian contribution to the DBI action.
As already stated, it is technically difficult to
consistently include the effect, especially including the
back-reaction to the D8-brane configuration.
Therefore in the paper, we instead took
only its energy into the system as the D4-brane action on the internal sphere.
Though there has been a lot of literature on holographic QCD with
finite baryon density, complete treatment of the non-Abelian instantons
seems to be still an open problem.
It is therefore very important to overcome this problem and to find a
consistent way to include the whole contribution in the future.

\textit{\bf Note added:}
During the preparation of the current version of the draft,
we have been informed that Prof.~Ooguri is also working on 
a similar problem about
the instability in the QGP phase.

\section*{Acknowledgments}
The authors thank Hong Liu, Shin Nakamura, Hirosi Ooguri, Dam Son and
Logan Wu for helpful discussions.
CPY thanks Hirosi Ooguri for the stimulating discussions and informing him of their
project on QGP phase.
WYC is supported by DOE grant DE-FG02-96ER40959. This work is also
supported by Taiwan's NSC grant 097-2811-M-003-012 and
97-2112-M-003-003-MY3. We also thank the support of NCTS.


\begin{thebibliography}{99}
\bibitem{Alford:1997zt}
 M.~G.~Alford, K.~Rajagopal and F.~Wilczek,
 ``QCD at finite baryon density: Nucleon droplets and color
 superconductivity,''
 Phys.\ Lett.\  B {\bf 422}, 247 (1998)
  [arXiv:hep-ph/9711395].
\\
M.~G.~Alford, K.~Rajagopal and F.~Wilczek,
``Color-flavor locking and chiral symmetry breaking in high density {QCD},''
  Nucl.\ Phys.\  B {\bf 537}, 443 (1999)
  [arXiv:hep-ph/9804403].


\bibitem{DGR}
  D.~V.~Deryagin, D.~Y.~Grigoriev and V.~A.~Rubakov,
 ``Standing wave ground state in high density, zero temperature QCD at large N(c),''
  Int.\ J.\ Mod.\ Phys.\  A {\bf 7} (1992) 659.

\bibitem{Shuster:1999tn}
  E.~Shuster and D.~T.~Son,
  ``On finite-density {QCD} at large N(c),''
  Nucl.\ Phys.\  B {\bf 573}, 434 (2000),

\bibitem{Domokos:2007kt}
  S.~K.~Domokos and J.~A.~Harvey,
  ``Baryon number-induced Chern-Simons couplings of vector and axial-vector
  mesons in holographic QCD,''
  Phys.\ Rev.\ Lett.\  {\bf 99}, 141602 (2007),
  [arXiv:0704.1604 [hep-ph]].

\bibitem{OO}
S.~Nakamura, H.~Ooguri and C.~S.~Park,
``Gravity Dual of Spatially Modulated Phase,''
Phys.\ Rev.\ {\bf D 81}, 044018 (2010),
arXiv:0911.0679 [hep-th].


\bibitem{SS1}
T.~Sakai and S.~Sugimoto,
``Low Energy Hadron Physics in Holographic QCD,''
Prog.\ Theor.\ Phys.\  {\bf 113} (2005) 843,
[arXiv:hep-th/0412141].

\bibitem{SS2}
T.~Sakai and S.~Sugimoto,
``More on a Holographic Dual of QCD,''
Prog.\ Theor.\ Phys.\  {\bf 114} (2005) 1083,
[arXiv:hep-th/0507073].

\bibitem{Aharony:2006da}
  O.~Aharony, J.~Sonnenschein and S.~Yankielowicz,
  ``A holographic model of deconfinement and chiral symmetry restoration,''
  Annals Phys.\  {\bf 322}, 1420 (2007),
  [arXiv:hep-th/0604161].
  


\bibitem{WittenBaryon}
E.~Witten,
``Baryons and Branes in Anti De Sitter Space,''
JHEP {\bf 9807} (1998) 006,
[arXiv:hep-th/9805112].

\bibitem{Hata:2007mb}
  H.~Hata, T.~Sakai, S.~Sugimoto and S.~Yamato,
  ``Baryons from instantons in holographic QCD,''
  Prog.\ Theor.\ Phys.\  {\bf 117} (2007) 1157
  [arXiv:hep-th/0701280].

\bibitem{Hong:2007kx}
  D.~K.~Hong, M.~Rho, H.~U.~Yee and P.~Yi,
  Phys.\ Rev.\  D {\bf 76}, 061901 (2007)
  [arXiv:hep-th/0701276].
  D.~K.~Hong, M.~Rho, H.~U.~Yee and P.~Yi,
  JHEP {\bf 0709}, 063 (2007)
  [arXiv:0705.2632 [hep-th]].

\bibitem{Witten:1998zw}
  E.~Witten,
  ``Anti-de Sitter space, thermal phase transition, and confinement in  gauge
 theories,''
  Adv.\ Theor.\ Math.\ Phys.\  {\bf 2}, 505 (1998),
  [arXiv:hep-th/9803131].

\bibitem{Bergman}
O.~Bergman, G.~Lifschytz and M.~Lippert,
``Holographic Nuclear Physics,''
JHEP {\bf 0711}, 056 (2007),
[arXiv:0708.0326 [hep-th]].

\bibitem{Lin}
F.~L.~Lin and S.~Y.~Wu,
``Holographic QCD with Topologically Charged Domain-Wall/Membranes,''
JHEP {\bf 0809}, 046 (2008),
[arXiv:0805.2933 [hep-th]].


\bibitem{Kim:2007zm}
  K.~Y.~Kim, S.~J.~Sin and I.~Zahed,
  ``The Chiral Model of Sakai-Sugimoto at Finite Baryon Density,''
  JHEP {\bf 0801} (2008) 002
  [arXiv:0708.1469 [hep-th]].

\bibitem{Gibbons:1976ue}
  G.~W.~Gibbons and S.~W.~Hawking,
  ``Action Integrals And Partition Functions In Quantum Gravity,''
  Phys.\ Rev.\  D {\bf 15}, 2752 (1977).


\bibitem{unstable}
  S.~Kobayashi, D.~Mateos, S.~Matsuura, R.~C.~Myers and R.~M.~Thomson,
  ``Holographic phase transitions at finite baryon density,''
  JHEP {\bf 0702}, 016 (2007)
  [arXiv:hep-th/0611099].
\\
  S.~Nakamura, Y.~Seo, S.~J.~Sin and K.~P.~Yogendran,
  ``Baryon-charge Chemical Potential in AdS/CFT,''
  Prog.\ Theor.\ Phys.\  {\bf 120}, 51 (2008)
  [arXiv:0708.2818 [hep-th]].
\\
  D.~Mateos, S.~Matsuura, R.~C.~Myers and R.~M.~Thomson,
  ``Holographic phase transitions at finite chemical potential,''
  JHEP {\bf 0711}, 085 (2007)
  [arXiv:0709.1225 [hep-th]].


\bibitem{Antonyan:2006vw}
  E.~Antonyan, J.~A.~Harvey, S.~Jensen and D.~Kutasov,
  ``NJL and QCD from string theory,''
  arXiv:hep-th/0604017.


  


\end{thebibliography}
\end{document}